\documentclass[aps,showpacs,twocolumn]{revtex4}
\usepackage{amsfonts}
\usepackage{amsmath}
\usepackage{amssymb}
\usepackage{hyperref}
\usepackage{graphicx}
\usepackage{dcolumn}
\usepackage{amsmath}

\renewcommand{\k}{\mathbf{k}}

\begin{document}

\title{Giant van Hove Density of States Singularities and Anomalies of Electron and Magnetic Properties in Cubic Lattices}

\author{P.A.~Igoshev and V.Yu.~Irkhin}
\affiliation
{620108 M.N. Mikheev Institute of Metal Physics,  Ekaterinburg, Russia,\\
620002 Ural Federal University, Yekaterinburg, Russia}
\begin{abstract}
Densities of states for simple (sc) and base-centered (bcc) cubic lattices with account of nearest and next-nearest neighbour hopping integrals $t$ and $t'$ are investigated in detail. It is shown that at values of $\tau \equiv t'/t = \tau_\ast$, corresponding to the change of isoenergetic surface topology, the formation of van Hove $\bf k$ lines takes place. At small deviation from these special values, the weakly dispersive spectrum in the vicinity of van Hove lines is replaced by a weak $\bf k$-dependence in the vicinity of few van Hove points which possess huge masses proportional to $|\tau - \tau_\ast|^{-1}$. The singular contributions to the density of states originating from van Hove points and lines are considered, as well as the change in the topology of isoenergetic surfaces in the $\bf k$-space with the variation of $\tau$. Closed analytical expressions for density of states as a function of energy and $\tau$ in terms of elliptic integrals, and power-law asymptotics at $\tau = \tau_\ast$ are obtained. Besides the case of sc lattice with small $\tau$ (maximum of density of states corresponds to energy level of X $\bf k$-point), maximal value of the density of states is always achieved at energies corresponding to \textit{inner} $\bf k$-points of the Brillouin zone positioned in high-symmetry directions, and not at zone faces.  
\end{abstract}


\maketitle
\section{Introduction}
\begin{figure*}[htp]
\includegraphics[width=0.4\textwidth]{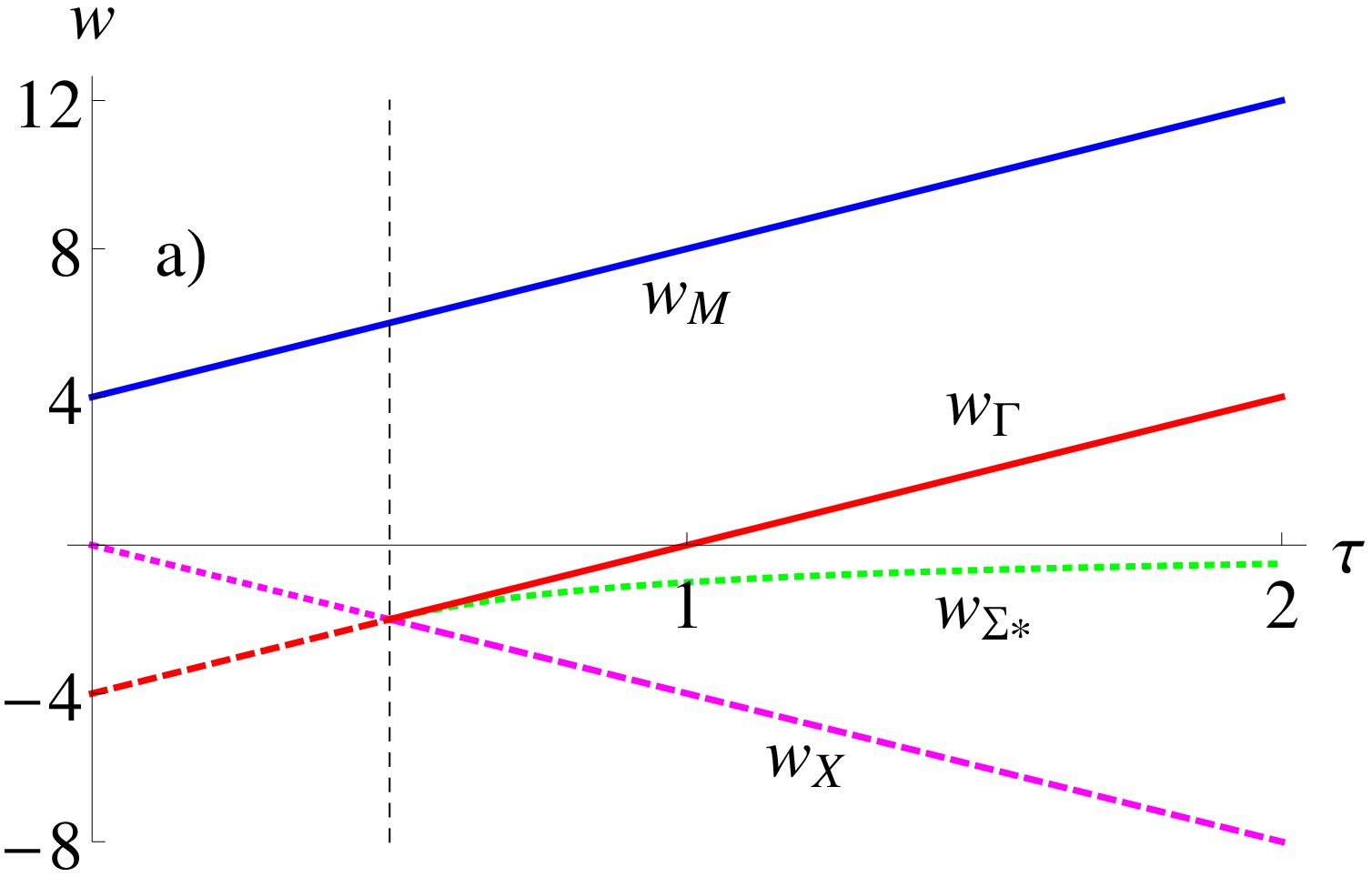}
\includegraphics[width=0.4\textwidth]{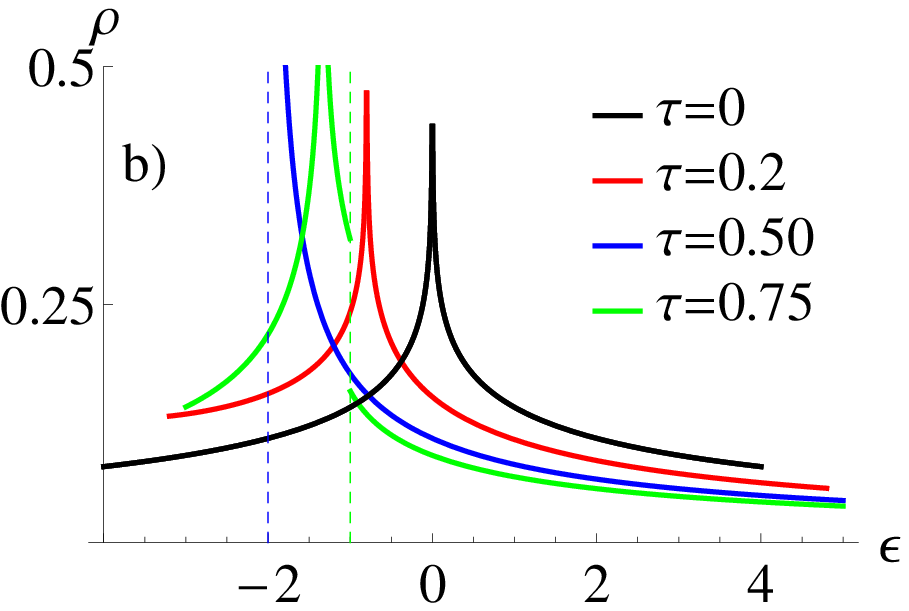}
\caption{\label{fig:square}(a) The dependence of important energy points on $\tau$ for the square lattice. Dashed (solid) line corresponds to local minimum (maximim), dotted line corresponds to saddle point. (b) DOS of the square lattice at different $\tau$. Dashed line shows the position of jump at $\epsilon = w_\Gamma$ for $\tau = 0.75$
}
\end{figure*}
The structure of the spectrum of non-interacting Bose and Fermi excitations is important for many problems of solid state physics. 
These include, for example, the electronic structure in the tight binding approximation and the spectrum of spin waves for free magnons~\cite{1969:Jelitto,1972:Swendsen}. 
Subsequently, these results can be used for the investigations of magnetic and electron properties within different microscopic models, e.g.~the Heisenberg, Kondo and Hubbard models with account of correlation effects. 
Besides that, peculiar properties of materials with density of states (DOS) singularities for electron and magnon excitations can be explained by the calculation of their first-principle spectrum. 
It is well known that DOS peculiarities near the Fermi level are crucial for ferromagnetic ordering, and large DOS values favor another Fermi-liquid instabilities (superconducting, antiferromagnetic etc.). 
\begin{figure*}[htp]
\includegraphics[width=0.4\textwidth]{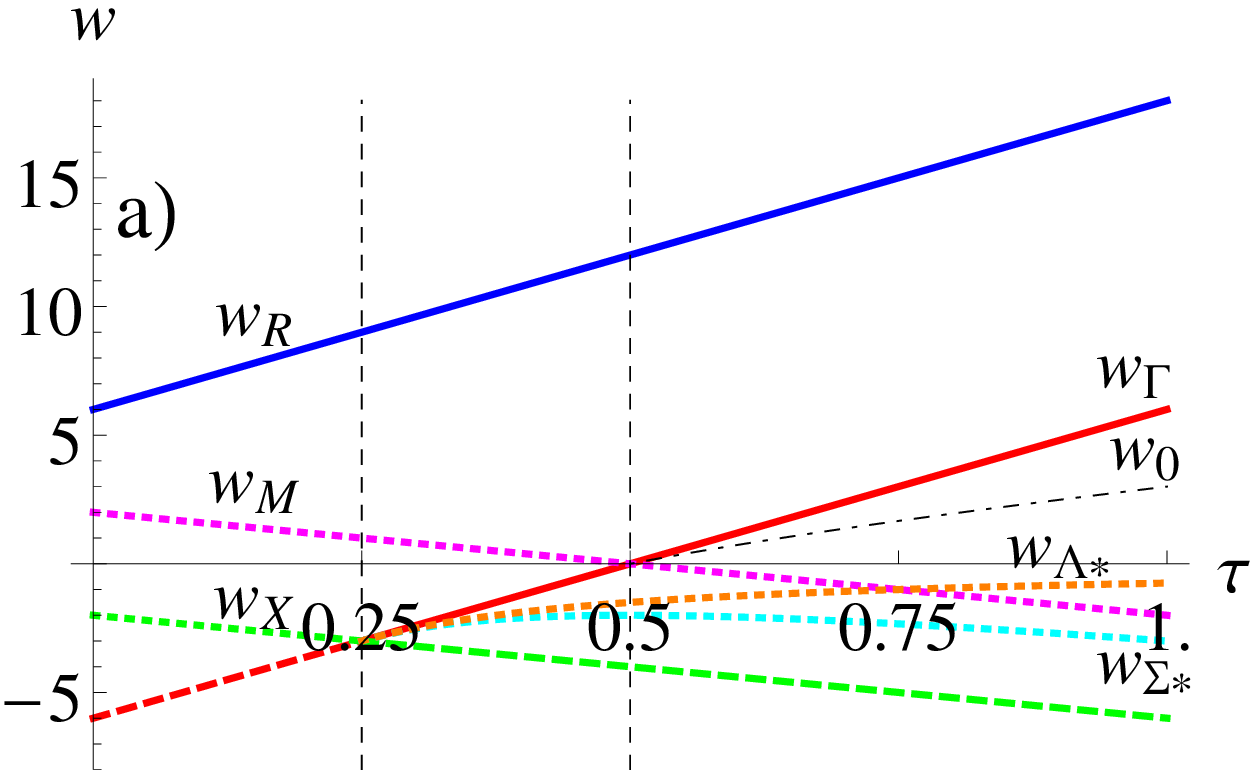}
\includegraphics[width=0.4\textwidth]{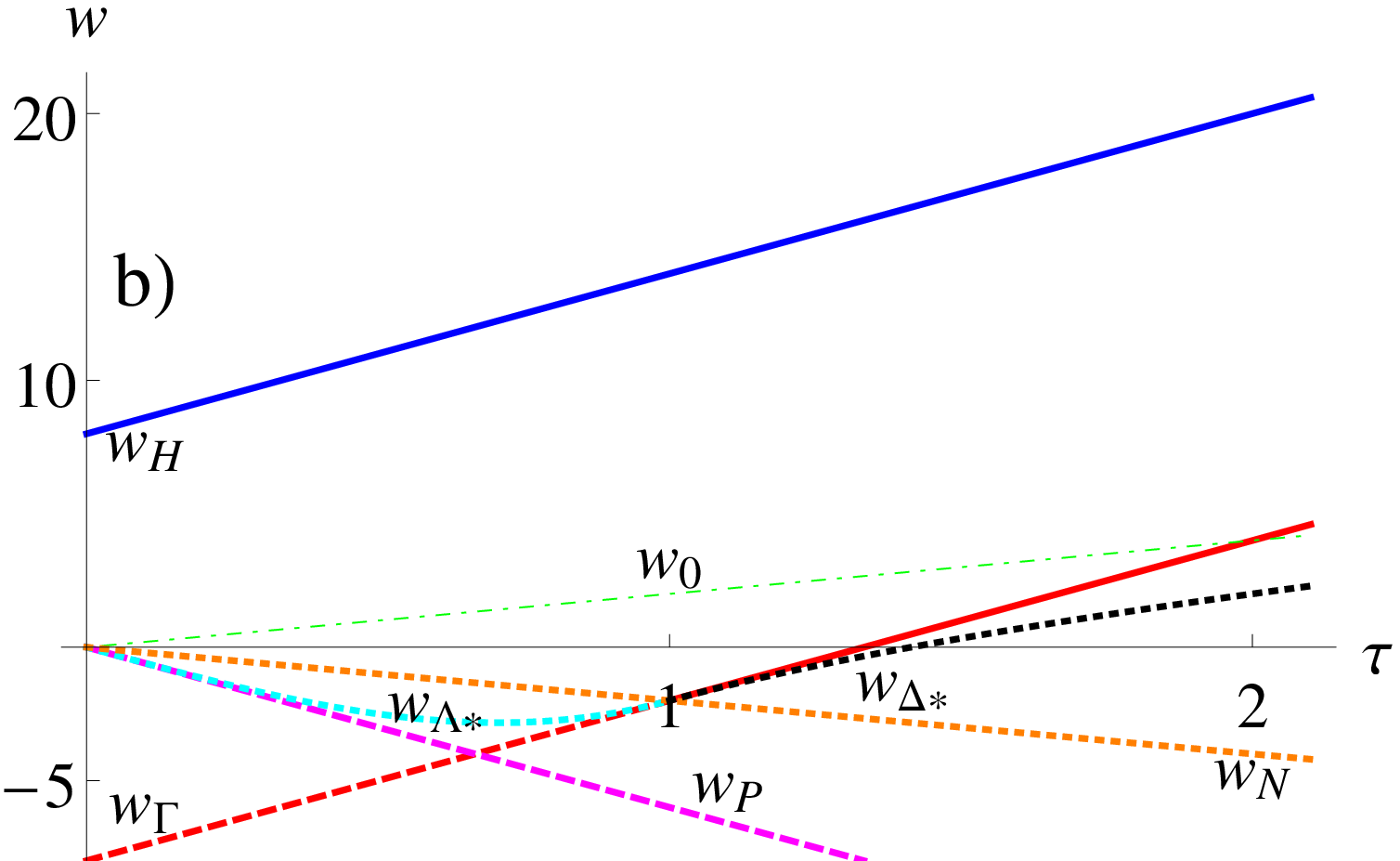}
\caption{$\tau$ dependence of van Hove levels $w$ for a) sc lattice, b) bcc lattice. The solid line corresponds to the local maximum of spectrum, the dotted line to the local minimum, the short-dotted line to the saddle point. 
\label{fig:w}
}
\end{figure*} 

Formation of narrow van Hove peaks of electron density of states results in the  localization  of the states and in an increase of correlation effects role. 
This fact was discussed in Ref.~\cite{1993:Trefilov} in the context of band structure calculations focusing on  anomalies of lattice and magnetic properties. 
A geometric origin  of electronic DOS peaks was investigated in detail in Ref.~\cite{1990:Peschanskih}. 
Simple examples are the bcc phase of Ca and fcc phase of Sr where weakly dispersive one-dimension manifolds (their prototype is a van Hove line) are positioned at the face of the Brillouin zone near the Fermi level.  
In bcc Ca these manifolds correspond to parts of P--N and N--H lines, whereas in fcc Sr to X--U, U--L, L--K, K--U, and K--W directions. 
An analysis shows that $D$-line of the bcc lattice typically looks like van Hove line in Li, V, Fe, Ba~\cite{1978:Moruzzy,1986:Papacostantopoulos}. 

For the paramagnetic phase spectrum of Ni, the L point of the Brillouin zone possesses a large effective mass and actually provides strong magnetic properties of Ni (``van Hove magnet'') and its temperature dependence of magnetic susceptibility~\cite{2017:Katanin}. 
The origin of weak ferromagnetism of ZrZn$_2$ is probably the contribution of a vicinity of the L point to the electron DOS, which was revealed in the first-principle calculation~\cite{2001:Santi}.
Large mass at the van Hove $\k$-point R provides weak ferromagnetism (with Curie temperature~41~K) of Ni$_3$Al~\cite{2011:Hamid}, which apparently invokes non-Fermi-liquid dependence of electron interaction contribution to resistivity at low temperatures due to the scattering by spin fluctuations~\cite{2005:Niklowitz}. 
Weak itinerant ferromagnet Y$_2$Ni$_7$ has low Curie temperature 58~K, small effective magnetic moment 0.44~$ \mu_{\rm B}$ and the Rhodes-Wohlfarth ratio about 6~\cite{Y2Ni7}. 
Band structure calculations predict the flat band spectrum in the vicinity of the van Hove point $\Gamma$ and weak dispersion in the P--Z direction of rhombohedral Brillouin zone~(possible van Hove line or a pair of large mass van Hove points)~\cite{2015:Singh}.

Strong peculiarities of DOS originate from a vicinity of closely positioned van Hove points~(we call this van Hove structure). An effect of such structure on electron and lattice properties, including many-body screening anomalies, is considered in Refs.~\cite{Katsnelson_Trefilov, 1985:Katsnelson_Trefilov, Katsnelson_Trefilov2,1992:Irkhin}.

Analytical expression for DOS and lattice Green's functions are obtained for a few two- and three-dimensional lattices within the nearest-neighbour hopping approximation which yields an even DOS function $\rho(\epsilon, \tau = 0)$ for bipartite lattices (square, sc and bcc)~\cite{1969:Jelitto,1971:Morita,1971:Katsura}.

However, for a number of modern solid state problems the next-nearest neighbour hopping has crucial value. 
Here belong the problem of electron spectrum of high-temperature superconductivity in copper oxides~($ t-J$ model)~\cite{1997:Izumov,2003:Damascelli} and metal-insulator transition in the antiferromagnetic phase~\cite{2016:Timirgazin,2016:Timirgazinfcc,2019:Igoshev}.

A conventional analysis of the density of electron states based on the band calculations within the framework of tetrahedron method~\cite{1994:Andersen} is unable to provide adequate description of van Hove singularities (which are crucial for magnetic and superconducting instabilities) because of incorrect account of contributions from small $\mathbf{v}(\mathbf{k})$ velocity points. 
An alternative for such an approach is an analytical approximation of the spectrum in the vicinity of high-symmetry points where the spectrum is almost flat (dispersionless)~\cite{2017:Stepanenko}. 
\begin{figure*}[htp]
\includegraphics[angle=-90,width=0.95\textwidth]{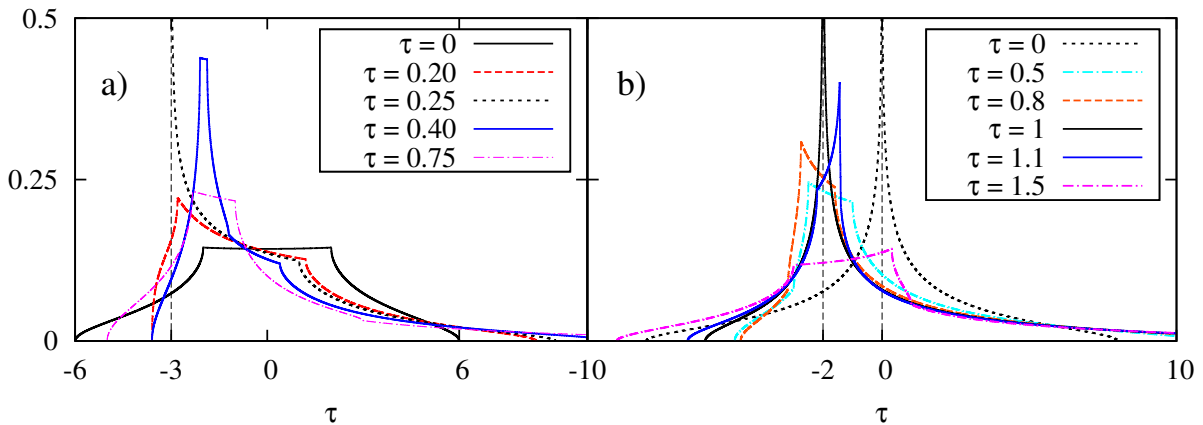}
\caption{DOS $t\rho(\epsilon)$ for sc (a) and bcc (b) lattice at different $\tau$. Vertical dashed line is positioned at the giant van Hove singularity energy levels. 
\label{fig:DOS_vs_e}
}
\end{figure*}

The van Hove singularities of one-particle excitation spectrum are promising for formation of novel or exotic instabilities. On the other hand, they are important for the long-standing problem of connection of magnetic properties of classical magnets like bcc-iron, cobalt etc. with their electron structure. 
Therefore, a corresponding general treatment of van Hove singularities for various three-dimensional lattices is useful; this is the aim of the present work.  

\section{Van Hove singularities for cubic lattices}

General definition of the density of states for any one-particle excitation spectrum $\varepsilon(\mathbf{k})$ reads 
\begin{equation}\label{eq:general_DOS_def}
	\rho(\epsilon) = \frac1{N}\sum_{\k} \delta(\epsilon - \varepsilon(\k)).
\end{equation}
\begin{widetext}
\center
\begin{tabular}{|c|c|c|c|}
\multicolumn{4}{l}{Table~1. Non-equivalent van Hove singularity $\k$-points for sc lattice, see~Fig.~\ref{fig:w}a.} \\
\multicolumn{4}{l}{Arrows denote the change of the type of van Hove singularity point as $\tau$ approaches $\tau^{\rm sc}_\ast$}from below.\\
\hline
$\mathbf{k}$ & $t(\mathbf{k})$ & \mbox{inverse masses} & \mbox{signature} \\
\hline
$\mathbf{k}_\Gamma$ & $w^{\rm sc}_\Gamma =  -6 + 12\tau$ & $a^{\rm sc}_1,a^{\rm sc}_1, a^{\rm sc}_1; a^{\rm sc}_1 = 2(1-4\tau)$&min$\rightarrow$max\\
\hline
$\mathbf{k}_{\rm R}$ & $w^{\rm sc}_{\rm R} = +6 + 12\tau$ & $-a^{\rm sc}_2,-a^{\rm sc}_2,-a^{\rm sc}_2; a^{\rm sc}_2 = 2(1+4\tau)$&max\\
\hline
$\mathbf{k}_{\rm X}$ & $w^{\rm sc}_{\rm X} = -2 - 4\tau$ &$2,2,2(4\tau - 1)$& $(++-)\rightarrow$min \\
\hline
$\mathbf{k}_{\rm M}$ & $w^{\rm sc}_{\rm M} = +2 - 4\tau$ & $-2,-2,2(4\tau + 1);$&  $(+--)$\\
\hline
$(0, p^{\rm sc}_1, p^{\rm sc}_1)$ & $w^{\rm sc}_{\varSigma\ast} = -\tau^{-1} + 2 - 4\tau$ & $-2\tau a^{\rm sc}_3, -a^{\rm sc}_3, a^{\rm sc}_3; a^{\rm sc}_3 = \tau^{-1} - 4$ &$\tau > 1/4$,$(++-)$  \\
\hline
$(p^{\rm sc}_2, p^{\rm sc}_2, p^{\rm sc}_2)$ & $w^{\rm sc}_{\varLambda\ast} = -3\tau^{-1}/4$& $2a^{\rm sc}_4,-a^{\rm sc}_4,-a^{\rm sc}_4; a^{\rm sc}_4 = (2\tau)^{-1}((4\tau)^2 - 1)$ &$\tau > 1/4$, $(+--)$ \\
\hline
\end{tabular}
\end{widetext}
Peculiarities of $\rho(\epsilon)$ result in anomalies of electron and magnetic properties, and corresponding phase transitions. 
There exists a general relation between peculiarities of DOS for the spectrum $\varepsilon(\k)$ and topological properties of the spectrum~\cite{1953:vanHove}. 
If excitation velocity $\mathbf{v}(\mathbf{k}) = \partial \varepsilon(\mathbf{k})/\partial\mathbf{k}$ vanishes at a $\mathbf{k}$ point, this is called the van Hove point. 
Energy levels $\epsilon = \varepsilon(\mathbf{k})$ corresponding to van Hove $\k$-points and only they provide peculiarities of DOS as a function of $\epsilon$.  
The type of the singularity corresponding to an isolated van Hove $\k$-point is determined by the signature of quadratic form with the matrix $\partial^2 \varepsilon(\mathbf{k})/\partial k_i\partial k_j$, specifying the behavior of the spectrum in the vicinity of this point.

For a three-dimensional lattice, a local $\k$-minimum (maximum) corresponds to one-sided  square-root DOS growth~(decrease) as $\epsilon$ moves away from the level $\epsilon_0 = \varepsilon(\k)$: $\rho(\epsilon) = \rho(\epsilon_0) + A\sqrt{\theta(\pm(\epsilon - \epsilon_0))|\epsilon - \epsilon_0|} + O(\epsilon - \epsilon_0)$  in the vicinity of $\epsilon_0$; the saddle $\k$-point of signature~$++-$~($+--$) corresponds to one-side square-root decrease $\rho(\epsilon) = \rho(\epsilon_0) - A\sqrt{\theta(\mp(\epsilon - \epsilon_0))|\epsilon - \epsilon_0|} + O(\epsilon - \epsilon_0)$,  $A > 0$ being a constant proportional to square-root of product of masses (more accurately, of eigenvalues of mass tensor).

If a spectrum, obtained within some first-principle calculations, possesses weak dispersion and is non-monotonous on a whole segment in some high-symmetry direction in the $\k$-space, there is one or a few $\k$-points with huge masses.

We assume that there is a hopping within first (integral $t$) and second (integral $t' = \tau t$) coordination spheres: $\varepsilon(\mathbf{k}) = t(\mathbf{k}; \tau)$. 
Here and below all energy scales are taken in $t$ units. 
For bipartite lattices the symmetry 
\begin{equation}\label{eq:DOS_inversion}
	\rho(-\epsilon; -\tau) = \rho(\epsilon; \tau)
\end{equation}
holds. 
Considering this property, we assume below $\tau \ge 0$.

\begin{widetext}
\center
\begin{tabular}{|c|c|c|c|}
\multicolumn{4}{l}{Table~2. Non-equivalent van Hove singularity $\k$-points for bcc lattice, see~Fig.~\ref{fig:w}b.} \\
\multicolumn{4}{l}{An arrow denotes the change of the type of van Hove singularity point at approaching $\tau$ from below to above $\tau^{\rm bcc}_\ast$}\\
\hline
$\mathbf{k}$ & $t(\mathbf{k})$ & \mbox{inverse masses} & \mbox{signature} \\
\hline
$\mathbf{k}_\Gamma$ & $w^{\rm bcc}_\Gamma = -8 + 6\tau$& $a^{\rm bcc}_1,a^{\rm bcc}_1,a^{\rm bcc}_1; a^{\rm bcc}_1 = 8(1-\tau)$& min$\rightarrow$max\\
\hline
$\mathbf{k}_{\rm H}$ & $w^{\rm bcc}_{\rm H} = +8 + 6\tau$ & $-a^{\rm bcc}_2,-a^{\rm bcc}_2,-a^{\rm bcc}_2; a^{\rm bcc}_2 = 8(1+\tau)$&max\\
\hline
$\mathbf{k}_{\rm P}$ & $w^{\rm bcc}_{\rm P} = -6\tau$ &$a^{\rm bcc}_3,a^{\rm bcc}_3,a^{\rm bcc}_3$; $a^{\rm bcc}_3 = 8\tau$& min \\
\hline
$\mathbf{k}_{\rm N}$ & $w^{\rm bcc}_{\rm N} = -2\tau$ &$-4\tau,4(1 + \tau),4(-1 + \tau)$& $(+--)\rightarrow(++-)$ \\
\hline
$(p^{\rm bcc}_1, p^{\rm bcc}_1, p^{\rm bcc}_1)$ & $w^{\rm bcc}_{\varLambda\ast} = 2\tau(2\tau^2 - 3)$ & $2a^{\rm bcc}_4,2a^{\rm bcc}_4,-a^{\rm bcc}_4; a^{\rm bcc}_4 = 8\tau(1-\tau^2)$& $\tau < 1$, $(++-)$\\
\hline
$(0, 0, p^{\rm bcc}_2)$ & $w^{\rm bcc}_{\Delta\ast} = 2\tau - 4\tau^{-1}$& $a^{\rm bcc}_5,a^{\rm bcc}_5,-a^{\rm bcc}_5; a^{\rm bcc}_5 = 8(\tau^{-1} - \tau)$ &$\tau > 1$, $(+--)$\\
\hline
\end{tabular}
\end{widetext}

For the square lattice, the van Hove saddle $\mathbf{k}$-point always exists by topological reasons and yields the logarithmic contribution to DOS. 
In the nearest-neighbour approximation for bipartite lattices, the singularity position at the center of the band coincides with the Fermi level for half-filling case, so that the ferromagnetic instability is suppressed by antiferromagnetic one (for which the free energy singular contribution is renormalized by correlation effects~\cite{2019:Igoshev}) owing to perfect nesting of the Fermi surface. 
Including the next-nearest hopping separates the logarithmic DOS peak from the Fermi level corresponding to half-filling and eventually results in the formation of the van Hove line $\Gamma$X (X$ = (0, \pi)$), $\varSigma$ direction, and in the change of the topology of the spectrum at $\tau = 1/2$~\cite{1997:Oles}. 
With further increase of $t'$, the local maximum of the spectrum is formed at the $\Gamma$ point, which yields a jump in DOS doubling its value, 
whereas the van Hove saddle point (denote it as $\varSigma^\ast$) migrates along the direction $\Gamma$M~(M being $(\pi, \pi)$) from the $\Gamma$ point (at $\tau < 1/2$ the saddle point is positioned at the X point)~\cite{2017:Igoshev},  see DOS plots at different $\tau$ in Fig~\ref{fig:square}b. 
Explicit expression for the square lattice DOS within the nearest and next-nearest neighbour hopping approximation has the form
\begin{multline}
        \label{eq:square_final_DOS_explicit}
	\rho_{\rm sq}(\epsilon; \tau) = \frac{2(1 + \theta(\psi(\epsilon, \tau)))}{\pi^2}\\
	\times\begin{cases}
		F\left(-\zeta(\epsilon, \tau),  \psi(\epsilon, \tau)\right), & w_{\rm X}(\tau) < \epsilon < w_{\rm r}(\tau), \\
		F\left(\zeta(\epsilon, \tau), \varphi(\epsilon, \tau)\right), & w_{\rm r}(\tau) < \epsilon < w_{\rm M}(\tau).		
	\end{cases},
\end{multline}
where $\theta(x)$ is the Heaviside step function and, here and below, we denote the energy level of $\k$ point A as $w_{\rm A}$,
\begin{eqnarray}
	w_{\rm r}(\tau) &=& \begin{cases}
	w_{\Gamma}(\tau), & \tau \le 1/2, \\
	w_{\varSigma^\ast}(\tau), & 1/2 < \tau.
	\end{cases}\\
\zeta(\epsilon, \tau) &=& 16 (1 + \tau\epsilon),\\ 
\varphi(\epsilon, \tau) &=& (\epsilon + 4\tau)^2,\\ 
\psi(\epsilon, \tau) &=& (4\tau - \epsilon)^2  - 16, 
\end{eqnarray}
\begin{eqnarray}\label{eq:sq.w_0}
	w_{\rm X}(\tau) = -4\tau,\\
	\label{eq:sq.w_s}
	w_{\varSigma\ast}(\tau) = -\tau^{-1}, \\
	\label{eq:sq.w_psi}
	w_{\Gamma,\rm M}(\tau) = \mp4 + 4\tau, 
\end{eqnarray}
see $\tau$-dependence of energy levels in Fig.~\ref{fig:square}a. The function 
\begin{equation}\label{eq:F_def}
	F(x, y) = \frac1{\sqrt{x}}\mathbb{K}(1 - y/x)
\end{equation}
is symmetric, $F(y, x) = F(x, y)$, singular when $x$ or $y$ tend to zero, and defined in the domain $x,y > 0$, $\mathbb{K}$ being the full elliptic integral of the first kind.

Thus, an especial important contribution to DOS originates from van Hove points which possess huge masses, or from merging of van Hove $\k$-points. The latter is typically connected with enhancement of one of masses, as it will demonstrated below.

DOS can be presented as the sum of three contributions
\begin{equation}\label{eq:DOS_exp}
	\rho(\epsilon; \tau) =  \mathcal{R}_\psi(\epsilon, \tau) + \mathcal{R}_{\varphi'}(\epsilon, \tau) + \mathcal{R}_\varphi(\epsilon, \tau).
\end{equation}

We obtained the explicit expressions for $\mathcal{R}_i$ for the sc lattice

\noindent
1. $\tau \le 1/4$. $\mathcal{R}^{\rm sc}_{\psi} = \mathcal{R}^{\rm sc}_{\varphi'} = 0$ and 
\begin{equation}\label{eq:R0_tau<=1/4}
	\mathcal{R}^{\rm sc}_\varphi =
	\begin{cases}
		\Phi_{\rm sc}(x^{\rm sc}_{\psi1}, +1),& w^{\rm sc}_\Gamma < \epsilon < w^{\rm sc}_{\rm X}(\tau),\\
		\Phi_{\rm sc}(-1, +1),& w^{\rm sc}_{\rm X}(\tau) < \epsilon < w^{\rm sc}_{\rm M}(\tau),\\
		\Phi_{\rm sc}(-1, x^{\rm sc}_{\psi2}),& w^{\rm sc}_{\rm M}(\tau) < \epsilon < w^{\rm sc}_{\rm R}.
	\end{cases}		
\end{equation}

2. $1/4 < \tau\le1/2$.
\begin{eqnarray}\label{eq:R1_sc}
	\mathcal{R}^{\rm sc}_\psi  &=&
	\begin{cases}
	2\Psi_{\rm sc}(x^{\rm sc}_{\varphi},+1),& w^{\rm sc}_{\rm X}<\epsilon < w^{\rm sc}_{\varSigma\ast},\\
	2\Psi_{\rm sc}(x^{\rm sc}_{\zeta -},x^{\rm sc}_{\zeta +}),& w^{\rm sc}_{\varSigma^\ast}< \epsilon < w^{\rm sc}_{\varLambda^\ast}.
	\end{cases};\\
\label{eq:R2_sc}
	\mathcal{R}^{\rm sc}_{\varphi'}  &=&
	\begin{cases}
	2[\Phi_{\rm sc}(x^{\rm sc}_{\psi1}, x^{\rm sc}_{\zeta -}) +\\+ \Phi_{\rm sc}(x^{\rm sc}_{\zeta +}, +1)], w^{\rm sc}_{\varSigma^\ast}<\epsilon < w^{\rm sc}_{\varLambda^\ast},\\
	2\Phi_{\rm sc}(x^{\rm sc}_{\psi1}, +1), w^{\rm sc}_{\varLambda^\ast}< \epsilon < w^{\rm sc}_\Gamma.
 	\end{cases};\\
\label{eq:R0_sc}
	\mathcal{R}^{\rm sc}_\varphi  &=&
	\begin{cases}	
	\Phi_{\rm sc}(-1,x^{\rm sc}_{\psi1}),& w^{\rm sc}_{\rm X}<\epsilon < w^{\rm sc}_\Gamma, \\
	\Phi_{\rm sc}(-1,+1),& w^{\rm sc}_\Gamma<\epsilon < w^{\rm sc}_{\rm M}, \\
	\Phi_{\rm sc}(-1,x^{\rm sc}_{\psi2}),& w^{\rm sc}_{\rm M}<\epsilon < w^{\rm sc}_{\rm R}.
	\end{cases}.
\end{eqnarray}	
3. $\tau>1/2$. Kinks of functions $\mathcal{R}^{\rm sc}_{\varphi}, \mathcal{R}^{\rm sc}_{\varphi'}$ at $\epsilon =  w_0^{\rm sc}(\tau) = 4\tau - \tau^{-1}$
cancel each other. 
\begin{eqnarray}\label{eq:R1_sc2}
	\mathcal{R}^{\rm sc}_\psi &=&	
	\begin{cases}
	2\Psi_{\rm sc}\left(x^{\rm sc}_{\varphi}, +1\right),& w^{\rm sc}_{\rm X}<\epsilon < w^{\rm sc}_{\varSigma^\ast},\\
	2\Psi_{\rm sc}\left(x^{\rm sc}_{\zeta -},x^{\rm sc}_{\zeta +}\right),& w^{\rm sc}_{\varSigma^\ast}<\epsilon <  w^{\rm sc}_{\varLambda^\ast}.
	\end{cases};\\
\label{eq:R2_sc2}
	\mathcal{R}^{\rm sc}_{\varphi'} &=&	
	\begin{cases}
	2\Phi_{\rm sc}\left(x^{\rm sc}_{\zeta +}, {\rm min}[x^{\rm sc}_{\psi2},+1]\right)\\ + 2\Phi_{\rm sc}\left(x^{\rm sc}_{\psi1}, x^{\rm sc}_{\zeta -}\right),  w^{\rm sc}_{\varSigma^\ast}<\epsilon <  w^{\rm sc}_{\varLambda^\ast},\\
	2\Phi_{\rm sc}\left(x^{\rm sc}_{\psi1},{\rm min}[x^{\rm sc}_{\psi2},+1]\right),  w^{\rm sc}_{\varLambda^\ast}<\epsilon <  w^{\rm sc}_0.
	\end{cases};\\
\label{eq:R0_sc2}
	\mathcal{R}^{\rm sc}_\varphi &=&
	\begin{cases}
	\Phi_{\rm sc}\left(-1, x^{\rm sc}_{\psi1}\right),& w^{\rm sc}_{\rm X}<\epsilon < w^{\rm sc}_{\rm M},\\
	\Phi_{\rm sc}\left(-1,x^{\rm sc}_{\psi1}\right) \\ \hspace{0.5cm}+ \Phi_{\rm sc}\left(x^{\rm sc}_{\psi2},+1\right),& w^{\rm sc}_{\rm M}< \epsilon <  w^{\rm sc}_0,\\
	\Phi_{\rm sc}\left(-1,x^{\rm sc}_{\psi2}\right) \\ \hspace{0.5cm}+ \Phi_{\rm sc}\left(x^{\rm sc}_{\psi1},+1\right),& w^{\rm sc}_0< \epsilon <  w^{\rm sc}_\Gamma,\\
	\Phi_{\rm sc}\left(-1,x^{\rm sc}_{\psi2}\right),& w^{\rm sc}_\Gamma<\epsilon < w^{\rm sc}_{\rm R}.
	\end{cases}.
\end{eqnarray}
Wihin these equations
\begin{equation}
\label{eq:SC:Psi_def}
	\Psi_{\rm sc}(x_1, x_2; \epsilon, \tau) = \frac2{\pi^3}\int\limits_{x_1}^{x_2}\frac{dx\,F(-\zeta_{\rm sc}(x; \epsilon, \tau), \psi_{\rm sc}(x; \epsilon, \tau) )}{\sqrt{1 - x^2}},
\end{equation}
\begin{equation}
\label{eq:BCC:Psi_def}
	\Phi_{\rm sc}(x_1, x_2; \epsilon, \tau) = \frac2{\pi^3}\int\limits_{x_1}^{x_2}\frac{dx\,F(\zeta_{\rm sc}(x; \epsilon, \tau), \varphi_{\rm sc}(x; \epsilon, \tau) )}{\sqrt{1 - x^2}},
\end{equation}
where
\begin{eqnarray}
	\zeta_{\rm sc}(x; \epsilon, \tau) &=&  16(\tau(\epsilon  + 2x) + (1 - 2\tau x)^2), \\
	\varphi_{\rm sc}(x; \epsilon, \tau) &=& (\epsilon + 2x + 4\tau)^2, \\
	\psi_{\rm sc}(x; \epsilon, \tau) & = & (\epsilon + 2x - 4\tau)^2 - 16(1 - 2\tau x)^2,
\end{eqnarray}
$x$ integration bounds have a form
\begin{eqnarray}
\label{eq:x12_sc_def}
	x^{\rm sc}_{\psi s}(\epsilon, \tau) &=& \frac12\frac{4((-1)^s + \tau) - \epsilon}{1 + 4(-1)^s\tau}, s = 1,2\\
\label{eq:xs_sc_def}	
	x^{\rm sc}_\varphi(\epsilon, \tau) &=& -2\tau - \epsilon/2,\\
\label{eq:xpm_sc_def}	
	x^{\rm sc}_{\zeta\pm}(\epsilon,\tau) &=& \frac{+\frac12 \pm \sqrt{-\frac34 - \tau \epsilon}}{2\tau}, \epsilon < w^{\rm sc}_{\varLambda^\ast}(\tau).
\end{eqnarray}

The topology of the spectrum is substantially changed with increasing $\tau$ above the value $\tau_\ast^{\rm sc} = 1/4$~(at which the $\k$-points form the whole $\varDelta$ line, the $\Gamma$ point changes its type from minimum to maximum, whereas the X point changes its type from saddle point to local minimum), see Fig.~\ref{fig:w}a and Table~1. 
\begin{figure*}
\includegraphics[angle=-90,width=0.9\textwidth]{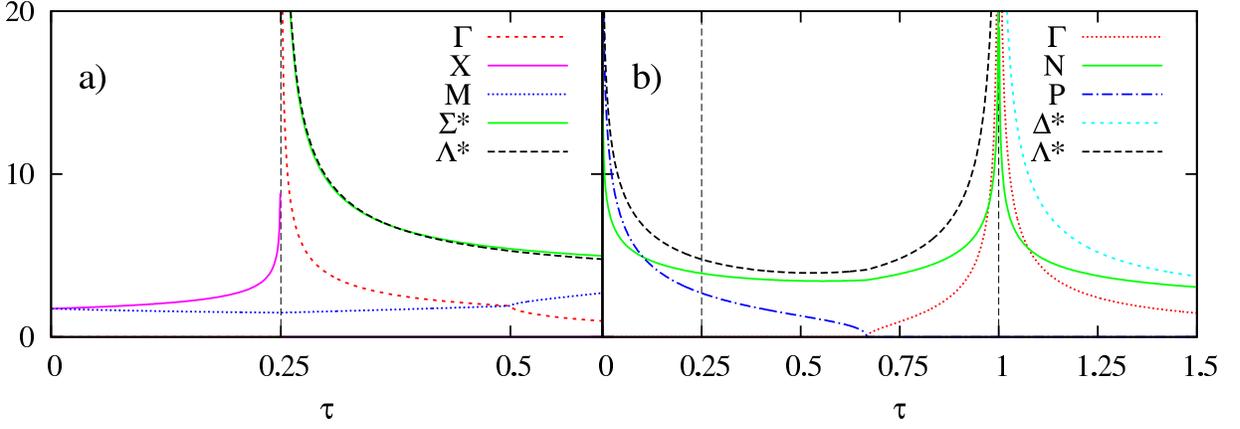}
\caption{$\tau$-dependence of  DOS, $W\rho(\epsilon;\tau)$,
at van Hove singularity energy levels for sc lattice (a), see Table~1 for the notations,  and bcc lattice(b), see Table~2, $W$ being bandwidth. For sc lattice, the values at $\epsilon = w^{\rm sc}_{\varLambda^\ast}$ and $w^{\rm sc}_{\varSigma^\ast}$ almost coincide, see the text. 
\label{fig:DOS_vs_tau}
}
\end{figure*}
At $\tau < \tau_\ast^{\rm sc}$  there exist only four van Hove points: minimum at $\Gamma$, maximum at R and two saddle points X, M with opposite signatures $(++-)$ and $(+--)$. Between corresponding pair of energy levels DOS plot forms a wide plateau.   
At $\tau > \tau_\ast^{\rm sc}$, two additional van Hove points $\varSigma^\ast$ (direction $\varSigma$) and $\varLambda^\ast$ (direction $\varLambda$) are split off from the $\Gamma$ point. Explicit expressions for these points and levels are presented in Table~1.  
During further increase of $\tau$, these points move away from the $\Gamma$ point and for each of them both three masses diverge like $|\tau-\tau_\ast^{\rm sc}|^{-1}$ as $\tau\rightarrow\tau_\ast^{\rm sc}$. 
Thus, at $\tau \gtrsim \tau_\ast^{\rm sc}$ van Hove line decays into van Hove struture (three closely positioned points $\Gamma, \varSigma^\ast$, $\varLambda^\ast$). 
The plot of $\rho(\epsilon, \tau)$ at $\tau > \tau_\ast^{\rm sc}$ forms a stable quasi-symmetric plateau between the energy levels $w^{\rm sc}_{\varSigma^\ast}, w^{\rm sc}_{\varDelta^\ast}$~(the signatures of van Hove points are analogous to ones of the pair X, M at $\tau < \tau_\ast^{\rm sc}$) with the height  $\propto(\tau - \tau_\ast^{\rm sc})^{-1/2}$, width at the top $4(\tau - \tau_\ast^{\rm sc})^2/\tau$ and universal dependence of DOS on the relative position of $\epsilon$ within the plateau top  (see Fig.~\ref{fig:DOS_vs_e}a,\ref{fig:DOS_vs_tau}a). 
The change of the DOS within the plateau is rather weak, which is related to simultaneous diverging of all three masses for both the $\k$-points $\varSigma^\ast$ and $\varLambda^\ast$. 
Maximal DOS value is achieved at $\epsilon = w^{\rm sc}_{\varSigma^\ast}$ or $w^{\rm sc}_{\varLambda^\ast}$, corresponding to inner van Hove $\mathbf{k}$ points of the Brillouin zone (see~Fig.~\ref{fig:DOS_vs_tau}a).

\begin{figure*}[htp]
\includegraphics[angle=-90,width=0.85\textwidth]{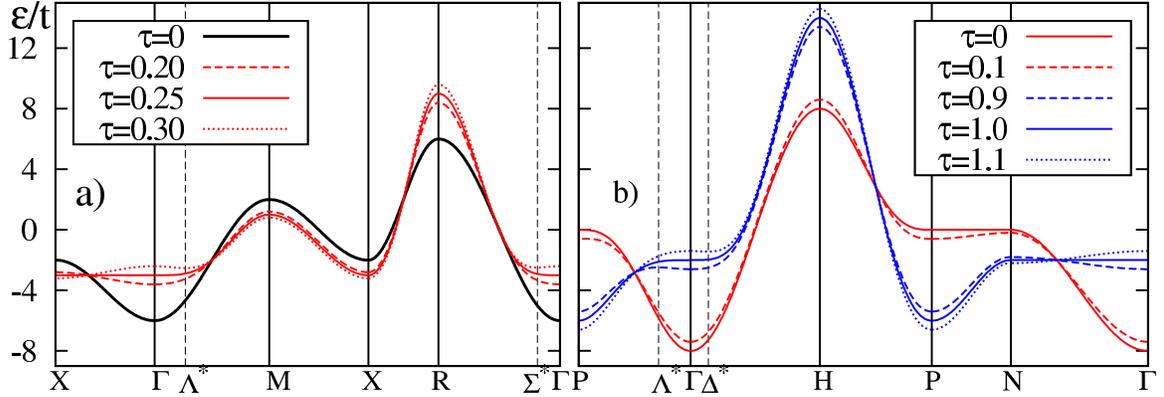}
\caption{The spectrum for sc (a) and bcc (b) lattices in high symmetry directions, the van Hove points being shown by vertical dashed line. 
\label{fig:ek}
}
\end{figure*} 

At $\tau = \tau_\ast^{\rm sc}$ the plateau contribution originating from above discussed three-point van Hove structure fully vanishes and the main contribution originates from the remaining background 
\begin{equation}\label{eq:SC:DOS_at_tau_ast}
\rho_{\rm sc}(-3 + \delta\epsilon, \tau_\ast^{\rm sc}) = \frac{B_{\rm sc}}{\sqrt[4]{\delta\epsilon}} + o(1),
\end{equation}
where $B_{\rm sc} = \frac{\sqrt{2}}{\pi}\frac{\Gamma(5/4)}{\Gamma^3(3/4)} = 0.222$. 
Thus, the giant van Hove singularity is formed originating from valleys of global minimum positioned on the $\varDelta$ line. 
An example of electron spectrum calculation for different $\tau$ in high-symmetry directions of Brillouin zone is shown in~Fig.~\ref{fig:ek}a.

For the bcc lattice we have the following explicit expressions

\noindent 1. $\tau \le 1$. 
\begin{eqnarray}\label{eq:R1_bcc_small_tau}
	\mathcal{R}^{\rm bcc}_\psi &=& \begin{cases}
	2\Psi_{\rm bcc}(0, x^{\rm bcc}_{\varphi}),& w^{\rm bcc}_{\rm P} < \epsilon < w^{\rm bcc}_{\varLambda^\ast},\\
	2\Psi_{\rm bcc}(0, x^{\rm bcc}_{\zeta}),& w^{\rm bcc}_{\varLambda^\ast} < \epsilon < w^{\rm bcc}_{\rm N}.
	\end{cases},\\
\label{eq:R2_bcc_small_tau}
	\mathcal{R}^{\rm bcc}_{\varphi'} &=& \begin{cases}
	2\Phi_{\rm bcc}(x^{\rm bcc}_{\zeta}, x^{\rm bcc}_{\psi1}),& w^{\rm bcc}_{\varLambda^\ast} < \epsilon < w^{\rm bcc}_{\rm N},\\
	2\Phi_{\rm bcc}(0, x^{\rm bcc}_{\psi1}),& w^{\rm bcc}_{\rm N} < \epsilon < w^{\rm bcc}_{0}.
	\end{cases},\\
\label{eq:R0_bcc_small_tau}
	\mathcal{R}^{\rm bcc}_\varphi &=& \Phi_{\rm bcc}(x^{\rm bcc}_{\psi1}, +1).
\end{eqnarray}
2. $\tau>1$. 
\begin{eqnarray}
\label{eq:R1_bcc_large_tau}
	\mathcal{R}^{\rm bcc}_\psi &=&
	\begin{cases}	
	2\Psi_{\rm bcc}(0, x^{\rm bcc}_\varphi),& w^{\rm bcc}_{\rm P} < \epsilon < w^{\rm bcc}_{\rm N},\\
	2\Psi_{\rm bcc}(x^{\rm bcc}_\zeta, +1),& w^{\rm bcc}_{\rm N} < \epsilon <  w^{\rm bcc}_{\varDelta^\ast}.	
	\end{cases},\\
\label{eq:R2_bcc_large_tau}
	\mathcal{R}^{\rm bcc}_{\varphi'} &=&
	\begin{cases}	
	2\Phi_{\rm bcc}(0, x^{\rm bcc}_{\zeta}),& w^{\rm bcc}_{\rm N} < \epsilon < w^{\rm bcc}_{\varDelta^\ast},\\
	2\Phi_{\rm bcc}(0, x^{\rm bcc}_{\psi1}) \\\hspace{0.5cm}+ 2\Phi_{\rm bcc}(x^{\rm bcc}_{\psi2}, +1),& w^{\rm bcc}_{\varDelta^\ast} < \epsilon < w^{\rm bcc}_\Gamma.	
	\end{cases},\\
\label{eq:R0_bcc_large_tau}
	\mathcal{R}^{\rm bcc}_\varphi &=&
\begin{cases}	
	\Phi_{\rm bcc}(x^{\rm bcc}_{\psi1}, x^{\rm bcc}_{\psi2}),& w^{\rm bcc}_{\varDelta^\ast} < \epsilon < w^{\rm bcc}_\Gamma,\\
	\Phi_{\rm bcc}(x^{\rm bcc}_{\psi1}, +1),& w^{\rm bcc}_\Gamma < \epsilon < w^{\rm bcc}_{\rm H},
\end{cases}
\end{eqnarray}
where the arguments $\epsilon, \tau$ are omitted for brevity. 
The kinks of the functions $\mathcal{R}^{\rm bcc}_{\varphi}, \mathcal{R}^{\rm bcc}_{\varphi'}$ at $\epsilon = w_0^{\rm bcc} = 2\tau$ cancel each other.
\begin{equation}\label{eq:Psi_bcc_def}
	\Psi_{\rm bcc}(x_1, x_2; \epsilon, \tau) = \frac2{\pi^3}\int\limits_{x_1}^{x_2}\frac{dxF(-\zeta_{\rm bcc}(x; \epsilon, \tau), \psi_{\rm bcc}(x; \epsilon, \tau) )}{\sqrt{x(1 - x)}},
\end{equation}
\begin{equation}\label{eq:Phi_bcc_def}
	\Phi_{\rm bcc}(x_1, x_2; \epsilon, \tau) = \frac2{\pi^3}\int\limits_{x_1}^{x_2}\frac{dxF(\zeta_{\rm bcc}(x; \epsilon, \tau), \varphi_{\rm bcc}(x; \epsilon, \tau) )}{\sqrt{x(1 - x)}},
\end{equation}
where
\begin{eqnarray}
	\zeta_{\rm bcc}(x; \epsilon, \tau) &=& 16(\tau(\epsilon  + 2\tau) + 4(1 - \tau^2)x), \\
	\varphi_{\rm bcc}(x; \epsilon, \tau) &=& (6\tau + \epsilon - 4\tau x)^2,\\
	\psi_{\rm bcc}(x; \epsilon, \tau) &=& (\epsilon - 2\tau -  4\tau x)^2 - 64x,
\end{eqnarray}
and $x$ integration bounds reads
\begin{eqnarray}\label{eq:x_bcc_bounds}
	x^{\rm bcc}_{\varphi}(\epsilon,\tau) &=& \frac{\epsilon + 6\tau}{4\tau}, w^{\rm bcc}_{\rm P} < \epsilon, \\
	x^{\rm bcc}_\zeta(\epsilon,\tau) &=& \frac{\tau(\epsilon + 2\tau)}{4(\tau^2 - 1)}, (\epsilon - w^{\rm bcc}_\zeta)(\tau - 1) > 0,\\
	x^{\rm bcc}_{\psi1,2}(\epsilon,\tau) &=& \frac{\left(\mp2 + \sqrt{\tau(\epsilon - 2\tau) + 4}\right)^2}{4\tau^2}, \epsilon > w^{\rm bcc}_{\varDelta^\ast}.
\end{eqnarray}

The topology of the spectrum is changed at $\tau^{\rm bcc}_{\ast} = 1$.
At $\tau < \tau^{\rm bcc}_{\ast}$ there are two van Hove $\k$-points of local minimum, $\Gamma$ and P, the van Hove $\k$-point H of local maximum and two saddle $\k$-points $\varLambda^\ast$ and N with opposite signatures, see~Table~2.  
At $\tau = 0$ the point $\varLambda^\ast$ coincides with P, and the levels $w^{\rm bcc}_{\rm N},w^{\rm bcc}_{\rm P}$ merge, so that the van Hove line $D$ is formed which produces the DOS singularity at the center of the band, $\rho_{\rm bcc}(\epsilon, \tau = 0) = (4/\pi^3)[\ln^2(64/|\epsilon|) - \pi^2/16] + o(\epsilon)$~\cite{1971:Katsura}. 
The cause of such energy dependence is linear vanishing of inverse transverse mass at the $D$ line. 
At $0 < \tau < \tau^{\rm bcc}_{\ast}$ the van Hove saddle $\k$-points  $\varLambda^\ast$ and N have opposite signatures, energy levels $w^{\rm bcc}_{\rm N},w^{\rm bcc}_{\varLambda^\ast}$ being boundaries of asymmetric plateau (for N point, only one mass diverges as $\tau \rightarrow 0,1$, whereas for $\varLambda^\ast$ three masses diverge and the right edge of plateau has a much larger DOS value), see Fig.~\ref{fig:DOS_vs_e}b. 
As $\tau$ increases van Hove point $\varLambda^\ast$ migrates from P point to $\Gamma$ point along $\varLambda$~line. 
When $\tau$ approaches $\tau^{\rm bcc}_{\ast}$, the width of the plateau vanishes and the energy levels $w^{\rm bcc}_{\rm N}$ and $w^{\rm bcc}_{\Gamma}$ merge, i.~e. the van~Hove line $\varSigma$ is formed. 

In the case $\tau > \tau^{\rm bcc}_{\ast}$ the N point signature changes~(see~Table~2), and the role of  $\varLambda^\ast$  point belongs to the $\varDelta^\ast$ point which has the same signature, splits off from the $\Gamma$ point at $\tau > \tau^{\rm bcc}_{\ast}$ and migrates along $\varDelta$ direction moving away from the point $\Gamma$. 
Similar to $\varLambda^\ast$ point at $\tau < \tau^{\rm bcc}_{\ast}$, for $\varDelta^\ast$ point all three masses diverge at $\tau\rightarrow\tau^{\rm bcc}_{\ast}$. 
This implies that at $\tau > \tau^{\rm bcc}_{\ast}$ an asymmetric plateau in $\rho_{\rm bcc}(\epsilon, \tau)$ plot is formed between the energy levels $w^{\rm bcc}_{\rm N},w^{\rm bcc}_{\varDelta^\ast}$ 
$\rho_{\rm bcc}(w^{\rm bcc}_{\varDelta^\ast},\tau)$ being much larger than 
$\rho_{\rm bcc}(w^{\rm bcc}_{\rm N},\tau)$. 
The energy level $w^{\rm bcc}_{\Gamma}$ is close to  $w^{\rm bcc}_{\varDelta^\ast}$, so that DOS value at $\epsilon = w^{\rm bcc}_{\Gamma}$ competes with a value at $\epsilon = w^{\rm bcc}_{\rm N}$, see.~Fig.~\ref{fig:DOS_vs_tau}b. 
The maximal DOS value is achieved at $\epsilon = w^{\rm bcc}_{\varLambda^\ast}$ for  $\tau < \tau^{\rm bcc}_{\ast}$  and at $\epsilon = w^{\rm bcc}_{\varDelta^\ast}$ for $\tau > \tau^{\rm bcc}_{\ast}$, both corresponding to inner $\k$-points of the Brillouin zone.

One can see the formation of spectrum extremums when $\tau$ deviates from values corresponding to van Hove lines formation ($\tau = 0$ and $\tau^{\rm bcc}_{\ast}$),~see~Fig.~\ref{fig:ek}b. 
At $\tau = \tau_\ast^{\rm bcc}$ DOS has the asymptotics
\begin{equation}
	\rho_{\rm bcc}(\epsilon = -2 + \delta\epsilon, 1) =
	-\frac{5}{2\pi^2} +
	\begin{cases}
		\frac{B^<_{\rm bcc}}{\sqrt[4]{-\delta\epsilon}},& \delta\epsilon < 0,\\
		\frac{B^>_{\rm bcc}}{\sqrt[4]{\delta\epsilon}},& \delta\epsilon > 0.
	\end{cases}  + o(1),
\end{equation}
where
\begin{eqnarray}
B^>_{\rm bcc} &=& \frac3{2\pi^{3}}
	\int\limits_{+1}^{+\infty}
	\frac{dt\mathbb{K}(1 - t^2)}{\sqrt{t}} = 0.33260,\\
B^<_{\rm bcc} &=& \frac1{\pi^3}\int\limits_0^{+\infty}\frac{dt}{\sqrt{t}}\mathbb{K}(-t^2) = 0.31358.
\end{eqnarray}

\section{Impact on thermodynamic properties}
As an example of van Hove singularity impact on thermodynamic properties, we consider the temperature dependence of specific heat. For free electrons, this quantity (per site and spin projection) reads 
\begin{equation}
	C(T) = -T^{-1}\int d\epsilon \rho(\epsilon)(\epsilon - E_{\rm F})^2f' (\epsilon),
\end{equation}
$f (\epsilon)$ being the Fermi function. When the Fermi level is well away from all peculiar DOS levels, at low temperature $C_{\rm reg}(T) \approx \gamma_{\rm reg} T$, where $\gamma_{\rm reg} = \pi^2\rho(E_{\rm F})/6 = \text{const}$. 
However, due to the presence of van Hove singularity, $\gamma(T) = C(T)/T$ can considerably deviate from constant.  

\begin{figure*}[htp]
\includegraphics[angle = -90,width=0.49\textwidth]{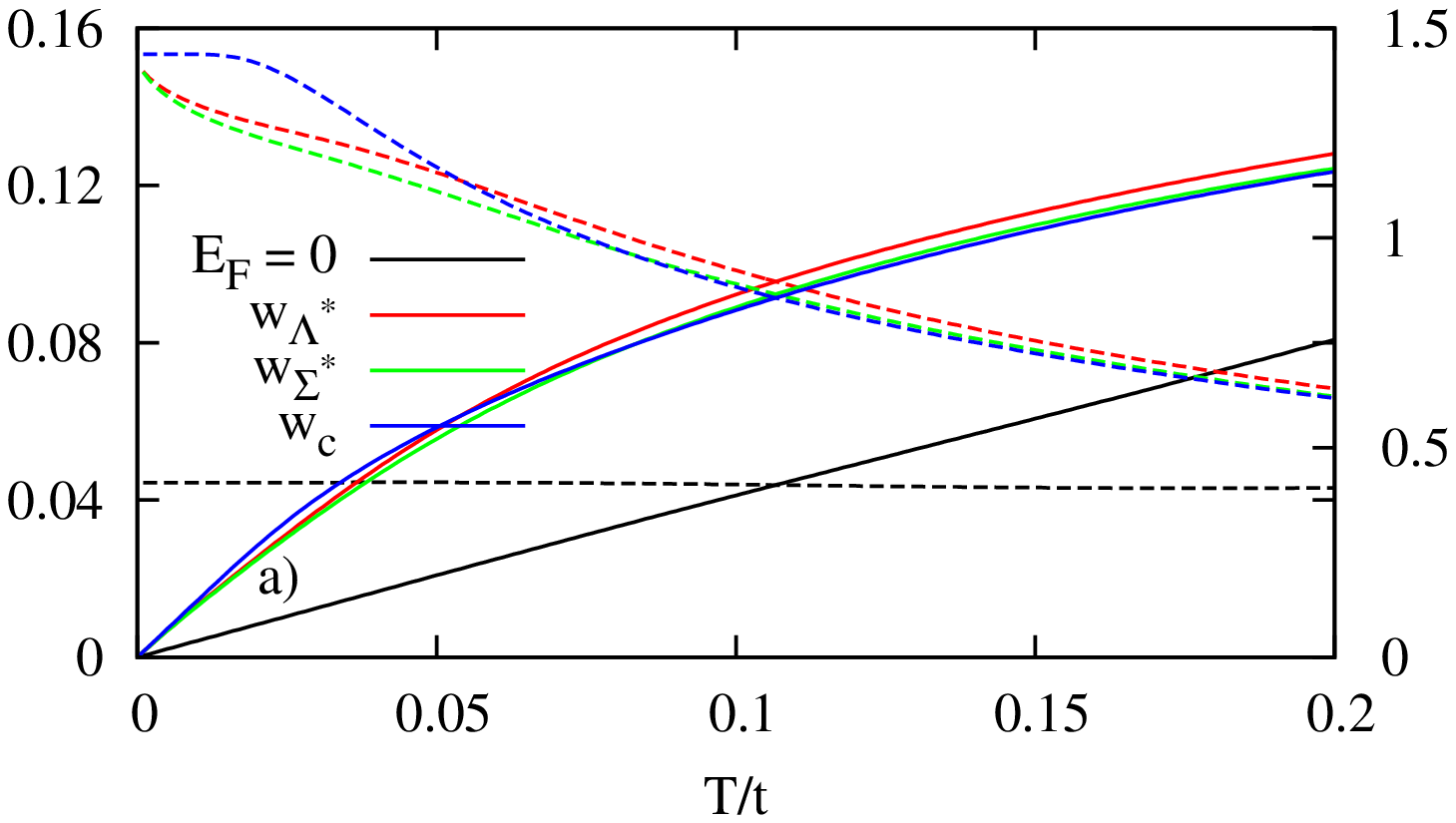}
\includegraphics[angle = -90,width=0.49\textwidth]{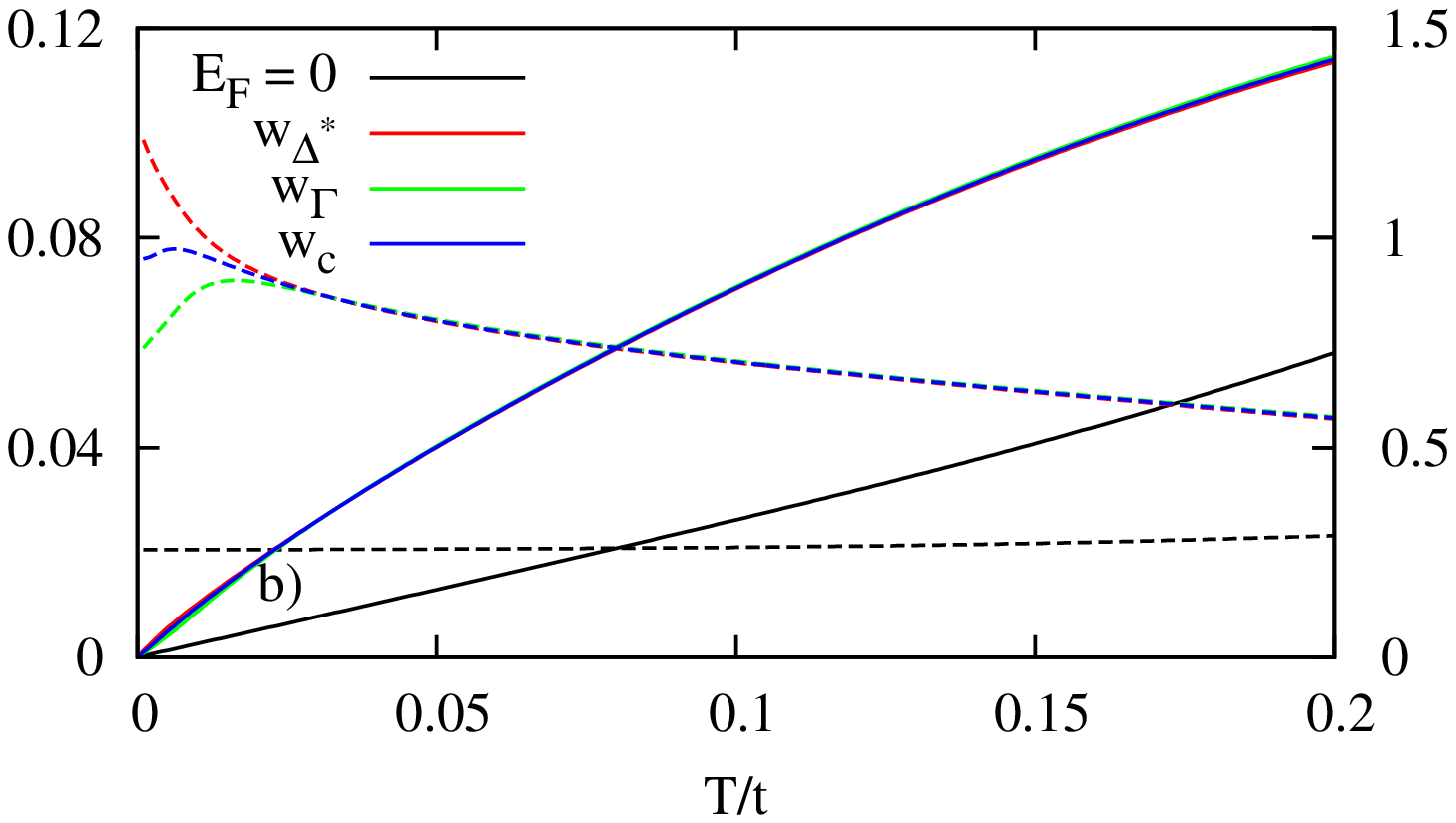}
\caption{\label{fig:C} Temperature dependences of specific heat $C$ (left axis, solid line) and $\gamma=C/T$ (right axis, dashed line) for (a) sc lattice, $\tau = 0.4$, the Fermi level being at quasi-symmetric $w_{\Sigma^\ast}$ -- $w_{\Lambda^\ast}$ plateau: $E_F = w_{\Sigma^\ast}, w_{\Lambda^\ast}, w_{\rm c} = (1/2)(w_{\Sigma^\ast} + w_{\Lambda^\ast})$; (b) bcc lattice, $\tau = 1.1$, the Fermi level being at non-symmetric $w_{\Delta^\ast}$ -- $w_{\Gamma}$ structure: $w_{\Delta^\ast}, w_{\Gamma}, w_{\rm c} = (1/2)(w_{\Delta^\ast} + w_{\Gamma})$. Lines of different colors correspond to different positions of $E_{\rm F}$; the regular case $E_{\rm F} = 0$ is shown for comparison.
}
\end{figure*}

We consider the case where the Fermi level $E_{\rm F}$ lies near the levels of van Hove structures for the sc lattice at $\tau>\tau^{\rm sc}_\ast$~(Fig.~\ref{fig:C}a) and bcc lattice at $\tau>\tau^{\rm bcc}_\ast$~(Fig.~\ref{fig:C}b). We choose the Fermi level position both at edges of the plateau and at its center. 

For the sc lattice, 
when $E_{\rm F}$ is on the top of the $\varSigma^\ast-\varLambda^\ast$ plateau and $\tau$ is rather close to $\tau_\ast^{\rm sc}$, the DOS value is large and practically insensitive to the change of $E_{\rm F}$ within the plateau. 
The position of $E_{\rm F}$ at the center of the plateau invokes the linear behavior of $C(T)$ at low temperature below some scale of order a plateau width (of order of $16|\tau-\tau_\ast^{\rm sc}|^2$), so that $\gamma_{\rm sc}(T)$ is large and almost constant, but rapidly falls down when the temperature increases above this scale. When $E_{\rm F}$ is at the left or right edge of the plateau the temperature behaviour is almost the same for both the cases, which manifests the symmetric form of the peak structure. 
At $\tau < \tau^{\rm sc}_\ast$ there are no peculiarities in the whole energy range. 

This implies that, unlike the case of weak ferromagnetism owing to the vicinity of large-mass $\k$-point, the ferromagnetism produced by the plateau is much more stable, possesses large Curie temperature and is expected to reproduce classical local moments picture of ferromagnetic ordering~\cite{Moriya}, see also~\cite{2015:Igoshev}. 

As the energy comes through the plateau boundaries, DOS decreases according a square-root law, so that the coefficient $A$ at the square root (see discussion above) is proportional to $(\tau - \tau_\ast^{\rm sc})^{-3/2}$.

For the bcc lattice at $\tau > \tau^{\rm bcc}_\ast$, the van Hove structure peak is formed by the close levels $\epsilon = w^{\rm bcc}_{\varDelta^\ast}$ and $w^{\rm bcc}_{\Gamma^\ast}$ corresponding to saddle~($\varDelta^\ast$) and maximum~($\Gamma$)~$\k$-points. In contrast to the sc lattice, this peak is typically more narrow and asymmetric (the value of DOS at $\epsilon = w^{\rm bcc}_{\varDelta^\ast}$ prevails). The narrowness of the peak causes merging $C(T)$ plots at different position of $E_{\rm F}$ as $T$ overcomes the energy scale of the order of peak width (of order of $4|\tau-\tau_\ast^{\rm bcc}|^2$). For $E_{\rm F} = w^{\rm bcc}_{\varDelta^\ast}$, $\gamma_{\rm bcc}$ rapidly falls down with increasing $T$ as temperature is below this energy scale. As the temperature overcomes the scale, the falling of $\gamma_{\rm bcc}(T)$ retains, but becomes weaker. As $E_{\rm F}$ 
deviates from $w^{\rm bcc}_{\varDelta^\ast}$, the dependence $\gamma_{\rm bcc}(T)$ becomes non-monotonous at low temperatures.
At $\tau < \tau^{\rm bcc}_\ast$ the level $\epsilon = w^{\rm bcc}_{\varLambda^\ast}$ plays the role of $w^{\rm bcc}_{\varDelta^\ast}$ at $\tau > \tau^{\rm bcc}_\ast$. 

We see that the temperature behaviour of the contributions from the van Hove singularities to thermodynamic properties is somewhat different for the sc and bcc lattices. 

The strong asymmetry of DOS with respect to $E_F$ can yield giant values of thermoelectric power when the Fermi level lies near the plateau boundaries, but beyond it. 

\section{Conclusions}
To conclude, we performed a systematic investigation of the effect of van Hove singularities on the spectrum of one-particle excitation and density of states for bipartite cubic lattices. We also considered in detail the dependence of spectrum topology and DOS behaviour on the next-nearest neighbour hopping integral.

For non-bipartite fcc lattice an analytic investigation of van Hove energy levels and numerical analysis of the DOS within the nearest and the next-nearest neighbour hopping approximation were performed only in a limited interval $\tau\in (-1, 0)$~\cite{1972:Swendsen} which did not allow to find van Hove singularity lines. 
However, at special value $\tau^{\rm fcc}_\ast = 1/2$ a giant van Hove line occurs at the bottom of the band $w^{\rm fcc}_{\rm b}$ producing DOS singularity of the type $ 1/\sqrt{\epsilon - w^{\rm fcc}_{\rm b}}$~\cite{1998:Ulmke}.  
The corresponding analysis of stability of this singularity and its influence on thermodynamic properties will be considered elsewhere.

We are grateful to M.I.~Katsnelson and A.O.~Anokhin for useful discussion. The research was carried out within the state assignment of Ministry of Science and Education of Russia (theme <<Quantum>> No.~AAAA-A18-118020190095-4) and with the support by Program 211 of the Government of the Russian Federation (Agreement 02.A03.21.0006).

\end{document}